\documentstyle[epsfig]{aipproc}

\begin{document}
\title{TeV Gamma Ray Emission from Southern Sky Objects and 
CANGAROO Project}

\author{T.Kifune$^{(1)}$, S.A.Dazeley$^{(2)}$, P.G.Edwards$^{(3)}$, 
T.Hara$^{(4)}$, Y.Hayami$^{(5)}$, S.Kamei$^{(5)}$, 
R.Kita$^{(6)}$, T.Konishi$^{(7)}$, A.Masaike$^{(8)}$, Y.Matsubara$^{(9)}$, 
Y.Matsuoka$^{(9)}$, Y.Mizumoto$^{(10)}$, M.Mori$^{(11)}$, 
H.Muraishi$^{(6)}$, Y.Muraki$^{(9)}$, T.Naito$^{(12)}$, K.Nishijima$^{(13)}$, 
S.Ogio$^{(5)}$, J.R.Patterson$^{(2)}$, 
M.D.Roberts$^{(1)}$, G.P.Rowell$^{(2)}$, T.Sako$^{(9)}$, K.Sakurazawa$^{(5)}$,
 R.Susukita$^{(14)}$, A.Suzuki$^{(7)}$, R.Suzuki$^{(5)}$,  
T.Tamura$^{(15)}$, T.Tanimori$^{(5)}$, G.J.Thornton$^{(2)}$, 
S.Yanagita$^{(6)}$, T.Yoshida$^{(6)}$, and T.Yoshikoshi$^{(1)}$}

\address{$^{(1)}$Institute for Cosmic Ray Research, University of Tokyo, 
Tanashi, Tokyo 188, Japan\\
$^{(2)}$Department of Physics and Mathematical Physics, 
  University of Adelaide, South Australia 5005, Australia, 
$^{(3)}$Institute of Space and Astrophysical Science, Sagamihara 229, Japan,  
$^{(4)}$Faculty of Management Information, 
Yamanashi Gakuin University, Kofu 400, 
$^{(5)}$Department of Physics, Tokyo Institute of Technology, Tokyo 152, 
$^{(6)}$Department of Physics, Ibaraki University, Mito 310, 
$^{(7)}$Department of Physics, Kobe University, Kobe 637,  
$^{(8)}$Department of Physics, Kyoto University, Kyoto 606,  
$^{(9)}$Solar-Terrestrial Environmental Laboratory, Nagoya University, Nagoya 464-01, 
$^{(10)}$National Astronomical Observatory of Japan, Tokyo 181, 
$^{(11)}$Department of Physics, Miyagi University of Education, Sendai 980, 
$^{(12)}$Department of Earth and Planetary Physics, University of Tokyo, 
Tokyo 113,  
$^{(13)}$Department of Physics, Tokai University, Hiratsuka 259,  
$^{(14)}$Institute of Physical and Chemical Research, Wako 351-01  
$^{(15)}$Department of Engineering, Kanagawa University, Yokohama 221, Japan}

\maketitle

\begin{abstract}
We report recent results of the CANGAROO Collaboration on 
very high energy gamma ray emission from pulsars, their nebulae,
SNR and AGN in the southern sky. 
Observations are made in South Australia using the imaging 
technique of detecting atmospheric Cerenkov light from 
gamma rays higher than about 1~TeV. 
The detected  gamma rays are most likely produced by the inverse Compton 
process by electrons which also radiate synchrotron X-rays. 
Together with information from longer wavelengths, 
our results can be used to infer the strength of magnetic field 
in the emission region of gamma rays 
as well as the energy of the progenitor electrons.
A description of the CANGAROO project is also given, 
as well as details of the new telescope of 7~m diameter which is 
scheduled to be in operation within two years.  
\end{abstract}

\section*{Introduction}

A firm foundation for VHE (Very High Energy) gamma ray astronomy \cite{Felix97}
 was laid  
with the use of the imaging \v Cerenkov technique to 
detect a signal from the Crab by the {\sc whipple} group \cite{Weekes89}.  
Detection of VHE emission from PSR~B1706$-$44 
soon followed by the 3.8~m imaging \v Cerenkov telescope
of the {\sc cangaroo} (Collaboration of Australia and Nippon(Japan)
for a GAmma Ray Observatory in the Outback) Project, which commenced
operation in 1992.
Gamma rays from this pulsar had been 
discovered by {{\sc cgro egret} soon after its launch.  
The 3.8 m telescope, located in the southern hemisphere, has the 
advantage of being able to study many Galactic objects near the 
Galactic center, 
while a number of \v Cerenkov imaging telescopes 
that have commenced operation in the northern hemisphere 
concentrate mainly on observing the Crab nebula and nearby AGNs.     
  
\section*{The CANGAROO Project}

The 3.8~m diameter telescope \cite{Hara93} is located
near Woomera, South Australia (at 136$^{\circ}$E and 31$^{\circ}$S, 
160 m above sea level). 
The camera  consists of 256 photomultiplier tubes with 
a total field of view of about 3$^{\circ}$ diameter, 
enabling a fine angular resolution to construct images of 
\v Cerenkov photons. 
The threshold energy of detectable gamma rays from the zenith 
was estimated to be 1$\sim$2~TeV before November 1996. 
The recoating of the mirror of the telescope with aluminium has since
reduced the threshold energy by a factor of 2. 

Another telescope of 10~m diameter is scheduled to commence  
operation in 1998, to exploit the region of $\sim$ 100~GeV energies.  
A radio antenna design is used for the construction of the main body 
of the telescope, which has an alt-azimuth mount. 
The light collecting mirrors attached to the parabolic antenna 
are spherical with 80~cm diameter and made of 
carbon reinforced plastic material which has the merit of having less weight
than glass. 
We will start with a total area equivalent to a 7~m diameter mirror.  
The focal length is 8~m, 
and a camera consists of 512 photomultiplier tubes of 
Hamamatsu R4124 of 1~ns rise time. 
 The diameter of the tube is 13~mm with the use of  
light guide to collect photons across $0.12^{\circ}$ 
 per tube.

\section*{Results of CANGAROO Observation }

\subsection*{Galactic Objects}

The galactic objects {\sc cangaroo} has observed    
 are listed in Table~1. 
In addition to the Crab nebula \cite{Tanimori94}, 
positive evidence has been obtained on PSR~B1706-44 \cite{Kifune95} 
and the Vela pulsar nebula \cite{Yoshikoshi97}.  

\begin{table} [h t]
\caption{List of Galactic Objects of CANGAROO Observation}
\begin{center}
\begin{tabular}{ccccc}  
objects & integral flux  & threshold   & observation & references \\
        & ($10^{-12}$ cm${-2}$ s$^{-1}$) & (TeV) & time (hrs) &  \\ \hline 
Crab  & 0.8  &  7 & 61 & \cite{Tanimori94}, \cite{Sakura97} \\
PSR~B1706$-$44  & 8   & $\sim$1 & 84 & \cite{Kifune95} \\
Vela  &  2.9 (nebula) & 2.5 & 119 + (28)$^{(a)}$ & \cite{Yoshikoshi97} \\
 &  $<$1.4 (pulsar position)& 2.5 & 119 + (28)$^{(a)}$ & \cite{Yoshikoshi97} \\
PSR~B1055$-$52  & $<$0.95   & 2 & 69 & \cite{Susukita97}  \\
PSR~B1758$-$23 & $< 2$   & $\sim$1 & 48 & \cite{Mori95} \\ 
PSR~B1259$-$63  & $\sim 4$ ?  & $\sim$3 & 38 + ($>$51)$^{(a)}$ & \cite{Sako97} \\
PSR~B1509$-$58  & $< 2 $  & 2 & 50 + ($>$31)$^{(a)}$ & \cite{Sako97}  \\
SN 1006  & a hint of a signal  & $\sim$2 & 35 + ($>$30)$^{(a)}$ &  \\ \hline
  \multicolumn{5}{l}{(a) data taken in 1997 with analysis yet to be done} \\
\end{tabular}
\end{center}
\end{table}

\noindent \underbar {\bf Crab}:  \hskip5truemm
The Crab is seen at zenith angles of 53$^{\circ}$ -- 60$^{\circ}$ 
 near its culmination from the observation site in Australia. 
The low elevation causes an increase of threshold energy 
 but  with an increase of detection area
\cite{Tanimori94}, 
enabling us to  enjoy excellent  
sensitivity at $\sim 10$~TeV energies. 
The energy spectrum observed extends to as high as $\sim 50$~TeV 
without an apparent change of slope from the  power law spectrum 
in 300~GeV to $\sim 1$~TeV region \cite{Sakura97}. 

\noindent \underbar {\bf EGRET pulsars; PSR~B1706$-$44 and Vela pulsar}: 
\hskip5truemm
These two EGRET pulsars are found to be VHE 
gamma ray emitters, but there is 
no evidence of VHE emission from PSR~B1055$-$52 \cite{Susukita97}. 
The signals from PSR~B1706$-$44 \cite{Kifune95} and 
the Vela \cite{Yoshikoshi97} show no modulation with the pulsar spin period,   
which suggests 
that pulsar nebula is, like the Crab, responsible for the VHE emission. 
The ratio of the observed VHE to X-ray luminosity of the nebulae are 
larger than in the case of the Crab.
The regime of synchrotron and inverse Compton processes of  
relativistic electrons to radiate X and gamma rays then indicates 
that the magnetic field in the nebulae is about two orders of magnitude 
weaker than the Crab nebula. 
VHE gamma rays are spatially peaked as offset to the 
south-east direction from the Vela  
pulsar by about $0.13^{\circ}$, and 
the spatial size of emission appears broader than the point spread function. 
The contribution from the pulsar position is given 
as the upper limit from the pulsar position in Table 1. 
We expect that our 1997 data with a reduced gamma ray threshold energy 
will be useful to infer the VHE spectrum. 
More knowledge with better accuracy,
particularly on the energy spectrum both in X-ray and VHE bands,
are required to restrict parameters of emission models  and 
to infer the structure of the nebulae.  

\noindent \underbar {\bf Other sources}: \hskip5truemm
Observations have not been made yet     
for objects such as young, short period pulsars or X-ray binaries, 
which are not detected by {\sc egret} so far but    
on which extensive VHE efforts have been made \cite{Kifune96}.  

Among the binary pulsars, PSR~B1259$-$63 is a peculiar object  
that has a highly eccentric orbital motion 
around a giant companion Be star. 
A VHE signal of $\sim 4\sigma$ was detected near the time of 
the previous periastron in 1994, and confirmation will be sought
in the data from around the next periastron in May 1997.
  
Supernova remnants are another target 
of prime importance for VHE gamma ray study.
One of the unidentified {\sc egret} sources is coincident with
the SNR W28 which is possibly associated with PSR~B1757$-$24. 
Evidence for point source emission was not detected 
for either W28 or PSR~B1758$-$23 \cite{Mori95}, 
though we can not exclude VHE emission, 
as appears in the earlier data \cite{Kifune93}, 
from the vicinity of the center of {\sc egret} error circle 
which is apparently shifted from W28 
towards the giant molecular cloud M20.    
Searches remain necessary for the unknown position of VHE emission 
possibly extended in the complex system 
of SNR, molecular cloud and pulsar. 
 PSR~B1509$-$58 is also embedded in a complex structure of plerion activity, 
and a considerable amount of data has been accumulated. 
An opportunity   
for VHE gamma rays to show direct evidence of 
shock acceleration of relativistic particles at the supernova shell 
is provided by shell type supernovae, such as SN1006 from which 
the {\sc asca} satellite detected 
non-thermal X-rays to suggest progenitor electrons 
up to $\sim 100$~TeV \cite{Koyama95}.  

\subsection*{Active Galactic Nuclei}

No TeV signal has been detected  
from nearby southern AGNs; 
Cen A; PKS 0521$-$36 (z=0.055); PKS 2316$-$42 (z=0.055); 
PKS 2005$-$49 (z=0.071); EXO 0423$-$08 (z=0.039) etc.\cite{Roberts97}.  
The upper limits are 
around $\sim 1 \times 10^{-12}$ cm$^{-2}$ s$^{-1}$ at 2~TeV, 
which lies near the quiescent Whipple flux levels of 
the northern BL~Lac objects which are VHE sources, {\it i.e.}, Mrk 421 and 501.

\section*{Discussions and Summary}

\begin{figure} 
\centerline{\epsfig{file=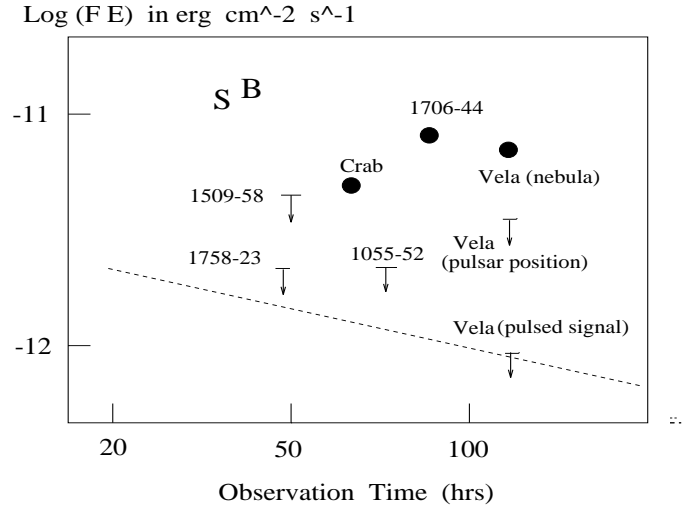,height=2.65in,width=3.5in}}
\vspace{10pt}
\caption{Energy flux observed by CANGAROO.}
\label{fig1}
\end{figure}

The {\sc cangaroo} detections 
are summarized in Fig.~1. 
The vertical axis is the product F$\cdot$E 
(in the unit of erg cm$^{-2}$ s$^{-1}$) of integral photon flux F  
and threshold energy E, 
and the horizontal axis the observation time used for analysis.
The detection sensitivity is near  
$10^{-12}$ erg cm$^{-2}$ s$^{-1}$ 
which corresponds to a luminosity of 
$10^{32}$ erg s$^{-1}$ at a distance of 1~kpc.
The significance of detection limited by statistical fluctuations 
is shown by the dotted line (Fig.~1),   
which is normalized to the detection
of 100 gamma rays at 1~TeV over 20~hrs.   
The preliminary result on PSR~B1259$-$63 is marked by ``B''. 
``S'' indicates the flux due to the inverse Compton process
 expected from SN1006  when the
magnetic field is as weak as the interstellar value. 
Loose upper limits on PSR~B1509$-$58 and on the Vela pulsar position 
are given due to a limited sensitivity of the current technique  
of imaging \v Cerenkov photons
to spatially extended emission. 

VHE gamma rays have a unique role to provide information about  
the high energy nature of these objects. 
There are an increasing number of interesting Galactic objects;
unidentified {\sc egret} sources,  pulsar nebulae and SNRs of 
non-thermal X-rays, objects with relativistic jets etc.,  
and  {\sc cangaroo} will continue to enjoy a $\lq$hard time'  
in choosing the targets for observation.

The project is supported by a Grant-in-Aid for Scientific Research 
of the Japan Ministry of Education, Science, Sports and Culture,  and 
also by the Australian Research Council.

\newpage


\begin{references}
\bibitem{Felix97}See review by Weekes, T.C. et al., in this  
{\it Proc. 4th CGRO Symp.} (1997).
\bibitem{Weekes89}Weekes, T.C. et al., {\it Ap. J.}\ {\bf
342}, 379 (1989).
\bibitem{Hara93}Hara, T. et al., {\it Nucl. Inst. Meth. Phys. Res.} 
{\bf 332}, 300 (1993).
\bibitem{Tanimori94}Tanimori, T. et al., {\it Ap. J. Lett} {\bf 429}, L61 
(1994).
\bibitem{Kifune95}Kifune, T. et al., {\it Ap. J. Lett} {\bf 438}, L91 
(1995).  
\bibitem{Yoshikoshi97}Yoshikoshi, T. et al., submitted to 
{\it Ap. \ J. Lett.} (1997).
\bibitem{Sakura97}Sakurazawa, K. et al., 
to appear in {\it Proc. 25th Int. Cosmic Ray Conf. (Durban)}, (1997).
\bibitem{Susukita97}Susukita, R., {\it Ph D Thesis, Kyoto University} (1997).
\bibitem{Mori95}Mori, M. et al., {\it Proc. 24th Int. Cosmic Ray Conf. 
(Rome)}, {\bf 2}, 491 (1995).  
\bibitem{Sako97}Sako, T. et al., to appear in {\it Proc. 25th Int. Cosmic Ray Conf. (Durban)}, (1997).  
\bibitem{Kifune93}Kifune, T. et al., {\it Proc. 23rd Int. Cosmic Ray Conf. 
(Calgary)}, {\bf 1}, 444 (1993).  
\bibitem{Kifune96}Kifune, T., {\it Space Science Review} {\bf 75}, 31 (1996). 
\bibitem{Roberts97}Roberts, M.D. et al., 
to appear in {\it Proc. 25th Int. Cosmic Ray Conf. (Durban)}, (1997).
\bibitem{Koyama95}Koyama, K. et al., {\it Nature} {\bf 378}, 255 
(1995). 
\end{references}
\end{document}